\documentclass[aps,pra,twocolumn,groupedaddress,superscriptaddress,showpacs]{revtex4}
\usepackage{natbib}
\usepackage[latin1]{inputenc}
\usepackage[dvips]{graphicx}
\usepackage{amsmath,amsbsy,amsfonts,amssymb}
\usepackage{bm}
\usepackage{textcomp}
\usepackage{multirow}

\begin{document}
\title{The states 1$^1\Sigma^+_u$, 1$^1\Pi_u$ and 2$^1\Sigma^+_u$ of Sr$_2$ studied by Fourier-transform spectroscopy}
\author{Alexander Stein}
\affiliation{Institut f\"ur Quantenoptik, Leibniz Universit\"at Hannover,
Welfengarten 1, D-30167 Hannover, Germany}
\author{Horst Kn\"ockel}
\affiliation{Institut f\"ur Quantenoptik, Leibniz Universit\"at Hannover,
Welfengarten 1, D-30167 Hannover, Germany}
\author{Eberhard Tiemann}
\affiliation{Institut f\"ur Quantenoptik, Leibniz Universit\"at Hannover,
Welfengarten 1, D-30167 Hannover, Germany}
\date{\today}
\begin{abstract}
A high resolution study of the electronic states $1^1\Sigma^+_u$ and $1^1\Pi_u$ which belong to the asymptote $4^1$D + $5^1$S and of the state $2(A)^1\Sigma^+_u$, which correlates to the asymptote $5^1$P + $5^1$S, is performed by Fourier-transform spectroscopy of fluorescence progressions induced by single frequency laser excitation. Precise descriptions of the potentials up to 2000 cm$^{-1}$ above the bottom are derived and compared to currently available \textit{ab initio} calculations. Especially for the state $1^1\Sigma^+_u$ large deviations are found. Rather weak and local perturbations are observed for the states $1^1\Pi_u$ and $2^1\Sigma^+_u$, while a strong coupling of the state $1^1\Sigma^+_u$ to the component $\Omega=0^+_u$ of the state $1^3\Pi_u$, which belongs to the asymptote $5^3$P + $5^1$S, is indicated.
\end{abstract}
\maketitle 

\section{Introduction}
\label{intro}

Our current spectroscopic work on the Sr$_2$ molecule is motivated by the interest in ultracold Sr$_2$ \cite{Zelevinsky2008,ZelevinskiC2008,KotochigovaProp2009}, where the desired information on the molecule for planning experiments was obtained up to now from \textit{ab initio} calculations \cite{Kotochigova2008}. In our earlier work on Ca$_2$ \cite{AllardCa2CoupledStates}, an isoelectronic molecule to Sr$_2$, we observed that the available \textit{ab initio} calculations are not as precise as they are e.g. for alkali molecules. Thus we want to test the accuracy of the theoretical calculations which are available for Sr$_2$ \cite{Kotochigova2008,Boutassetta,Czuchaj}.  

\begin{figure}
\resizebox{0.5\textwidth}{!}{%
\includegraphics{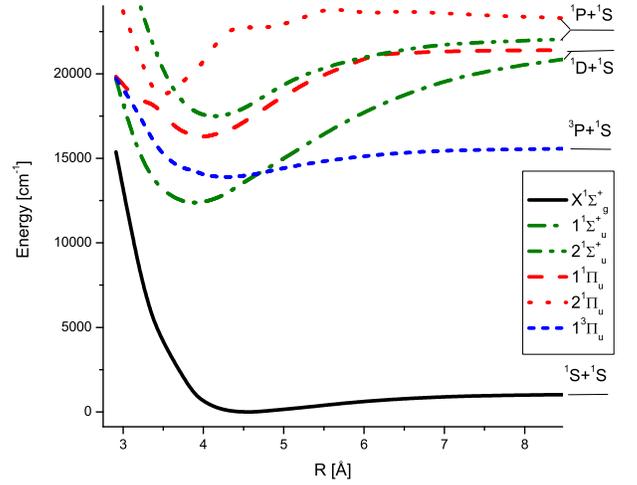}}
\caption{(Color online) Overview of the electronic states needed for interpretation in this paper. The potentials stem from the \textit{ab initio} calculations \cite{Boutassetta,Frecon_priv}.}
\label{Overview}
\end{figure}

In our previous publications on this molecule \cite{SteinSr2008,SteinSrXAsymptote} we mainly concentrated our effort on the ground state X$^1\Sigma^+_g$. For the present work we performed extensive new measurements to increase the knowledge on the excited states.
Figure \ref{Overview} gives an overview of the electronic states which are important in the current work. The potentials are taken from the \textit{ab initio} work \cite{Boutassetta,Frecon_priv}, which shows the closest approach to our present observations. 

We report on the successful observation of the state $1^1\Sigma^+_u$, which is presumably only known from the spectroscopy in rare gas matrices \cite{Miller1977,Miller1978,Miller1980}, on the observation of the state $1^1\Pi_u$, which, to our knowledge, was not experimentally known before, and on an extended data set of the state $2^1\Sigma^+_u$, which was in the past the mostly investigated excited state, called A state \cite{Bergemann1980,Gerber_Sr2_1984}, and was already used for exciting the fluorescence to the ground state in our first paper on this molecule \cite{SteinSr2008}.  

The paper is organized as follows: Section \ref{sec:exp} describes the experimental methods, Sec. \ref{sec:analysis} describes the obtained data set and discusses the measurement uncertainties, Sec. \ref{sec:results} presents the resulting sets of Dunham coefficients and potential descriptions and compares them to the currently available theoretical calculations, while Sec. \ref{sec:conclusion} gives conclusions and a short outlook.

\section{Experiment}
\label{sec:exp}

The experimental procedure and the apparatus used are the same as in \cite{SteinSr2008}. Strontium is filled into a stainless steel heatpipe, which is heated to a temperature of about 1220 K with 20 mbar of argon as a buffer gas. To keep the optical path free from condensing Strontium which grows as crystals at both ends of the heatpipe, the oven has to be moved every two hours to melt and bring the condensed Sr back to the heated zone. This means, that the heated cell is not working perfectly as a heatpipe where the condensed substance flows back to the heated region.

A Fourier-transform spectrometer of type Bruker IFS 120 HR resolves fluorescence progressions induced by a cw laser beam. For most spectra a resolution of 0.05 cm$^{-1}$ was chosen and typically about 20 scans of the spectrometer were averaged. This method is most efficient if fluorescence to the wanted state can be observed by the excitation of higher lying states, because each fluorescence progression contains typically information on a larger amount of different rovibrational levels (with $\Delta$J$\pm$1 or 0) of the lower state. However, for the excited state only one excited rovibrational level is involved. Fluorescence to states other than the ground state was searched without success. Even with a high power Ar$^+$ laser (2.5 W), which was used to investigate the existence of possible very weak transitions and two-photon excitations, no fluorescence to states other than the ground state was observed. This approach was also tried unsuccessfully with UV lines of that laser.

An alternative for a more efficient acquisition of a large amount of data could have been absorption methods. It was tried to record the absorption spectrum of Sr$_2$ with a white light source, but even with long averaging times (1000 scans which takes more than 30 hours) only signal-to-noise ratios of up to 5 could be reached. Moreover, the comparison with simulated spectra based on the potential regions already known from our earlier work \cite{SteinSr2008} showed that the visible absorption peaks do not stem from single strong lines but from an overlap of many lines. This finding is in good agreement with the recorded fluorescence spectra, where the laser, independent of its exact excitation frequency chosen within the frequency range of a strong electronic transition, excites each time more than a dozen of rovibrational levels. The reason for this is due to the large mass of the Sr dimer together with the high working temperatures implying that almost all existing rovibrational levels (about 5400 for the reference isotopologue $^{88}$Sr$_2$) of the relatively flat ground state potential (the binding energy is 1081.64(2) cm$^{-1}$ \cite{SteinSrXAsymptote}) are significantly thermally populated. Other methods as polarization labeling spectroscopy were already tried by the authors of \cite{Gerber_Sr2_1984} without success because of too weak molecular signals \cite{SchneiderDip}.

Due to the high amount of simultaneous excitations the method of recording fluorescence induced by the direct laser excitation of the wanted electronic states is not as inefficient as it seems to be at first glance. The recording of typical 20 scans with a resolution of 0.05 cm$^{-1}$ takes about 20 minutes yielding information on all simultaneously excited levels. In addition, the strongest progressions sometimes are accompanied by a few collisionally induced rotational satellites. During the recording process the experimentator can already use a widely automated software to assign the lines of the previously recorded spectrum and incorporade the obtained transition frequencies and level energies to improve the predictions for the next excitation frequencies. 

\subsection{Observation of the state 1$^1\Sigma^+_u$}

With a titanium sapphire laser in the range from 12000 cm$^{-1}$ to 14000 cm$^{-1}$ we collected a sufficient amount of data for a precise description of the bottom of the potential of the state $1^1\Sigma^+_u$. Only $v'=0$ for this state was not observed because of too small Franck-Condon overlap with the vibrational manifold of the ground state. During this investigation we also tried a dye laser operated with DCM dye, searching the frequency range from 15250 to 16700 cm$^{-1}$ for higher excited vibrational levels of this state. Nine progressions were observed which were excited close to 15250 cm$^{-1}$ and originate from three neighboring vibrational states of that electronic state. These levels with term energies around 15500 cm$^{-1}$ above the potential minimum of the ground state (but below the atom pair asymptote $^1$S$_0 + ^3$P$_1$ at 15586 cm$^{-1}$) show significant local perturbations which became obvious from plots of these term energies as a function of $J(J+1)$. Additionally, 18 progressions were observed with excitation frequencies around 16250 cm$^{-1}$ belonging to seven different vibrational levels. They form a structure of fragmentary bands in the term energy region between 16790 cm$^{-1}$ and 16890 cm$^{-1}$. The line intensities in the center of these bands are as strong as the strongest ones observed for low vibrational levels of $1^1\Sigma^+_u$. The observation of fragmentary bands clearly indicate predissociation to the $^3$P asymptotes between 15586 cm$^{-1}$ ($^1$S$_0 + ^3$P$_1$) and 15980 cm$^{-1}$ ($^1$S$_0 + ^3$P$_2$). Though the frequency interval from 16000 cm$^{-1}$ to 18500 cm$^{-1}$ was intensively investigated using dye lasers operated with DCM, Rhodamine 6G and Rhodamine 110, only three progressions belonging to levels outside the mentioned energy interval were observed and excited with laser frequencies close to 18250 cm$^{-1}$. These three levels do not fit into the now well known potential energy curves of the other two electronic states in this frequency region (1$^1\Pi_u$ and 2$^1\Sigma^+_u$) and the progressions have line intensities which fit to Franck-Condon factors expected for highly excited levels of the state 1$^1\Sigma^+_u$. Since also the observed rotational and vibrational spacings fit into the expectations for this state, it seems to be obvious that the levels belong to a second region with fragmentary bands of this state.

\subsection{Observation of the state 1$^1\Pi_u$}

During the investigation of the fragmentary bands of the state $1^1\Sigma^+_u$ one progression of Q lines was observed which was later assigned to $v'=1$ of the state $1^1\Pi_u$ correlating to the asymptote $^1$S$_0 + ^1$D$_2$, see Fig. \ref{Overview}. This state was further investigated, first by searching for neighboring rotational levels $J'$ of the same $v'$ and then by searching for neighboring $v'$ levels. For this search theoretical $B_e$ and $\omega_e$ from \cite{Boutassetta} were used until enough levels were available to fit a preliminary set of Dunham coefficients which then were used for further extrapolations. The effort in time for the investigation of the state $1^1\Pi_u$ was significantly higher than for the $^1\Sigma^+_u$ states because the observed line intensities of the Q progressions belonging to this state are typically a factor of seven lower than the line intensities of the P, R progressions from the states $^1\Sigma^+_u$. Because of this fact the chance of finding accidentally excited transitions is low and the precision needed for successful predictions of new transitions is high. Vibrational levels of the state $1^1\Pi_u$ were investigated up to $v'=22$ using dye lasers with DCM, Rhodamine 6G and Rhodamine 110 reaching excitation energies up to 18500 cm$^{-1}$. 

\subsection{Observation of the state 2(A)$^1\Sigma^+_u$}

Since the transition strengths from the state 2(A)$^1\Sigma^+_u$ showed to have on average a similar amplitude as for the state $1^1\Sigma^+_u$, it was also possible to collect a sufficient amount of data for a precise description of the potential up to 2000 cm$^{-1}$ above its bottom using dye lasers with Rhodamine 6G and Rhodamine 110. Nevertheless, the overwhelming amount of data obtained for this state is a side product of the investigation of the ground state \cite{SteinSr2008,SteinSrXAsymptote} and the intensive search for energy levels of the state $1^1\Pi_u$ as mentioned above. Using the visible lines of an Ar$^+$ laser a large amount of time was investigated to search for highly excited levels of this state, but at the end this was not very successful. A few progressions were found by exciting with the 514 nm and the 496 nm lines. To identify few weak progressions excited with the 488 nm line much longer averaging time and the use of a single-mode assembly for the Ar$^+$ laser were necessary. For the more blue lines of the laser no fluorescence progressions were found. This strong decrease of the fluorescence intensities towards higher excitation energies cannot be explained by low populations of those ground state levels with best Franck-Condon overlap since the ground state population changes only by a factor of two from the potential minimum to the asymptote for the high working temperature of 1220 K. Because of the high level density in the long range regions of both the excited and the ground state the number of simultaneous excitations strongly increases for high $v'$ and because the maximum total absorption of the Sr$_2$ molecules in the heatpipe is only of about 10\% the strength of the total fluorescence should strongly increase. But we observe that it stays roughly constant and the individual progressions become too weak for a reliable analysis. The most likely explanation is again predissociation which could be caused by a growing Franck-Condon overlap of the higher vibrational states $v'$ with the continuum levels of the lower lying triplet and singlet states.  
 
\section{Data set}
\label{sec:analysis}

\begin{figure}
\resizebox{0.5\textwidth}{!}{%
\includegraphics{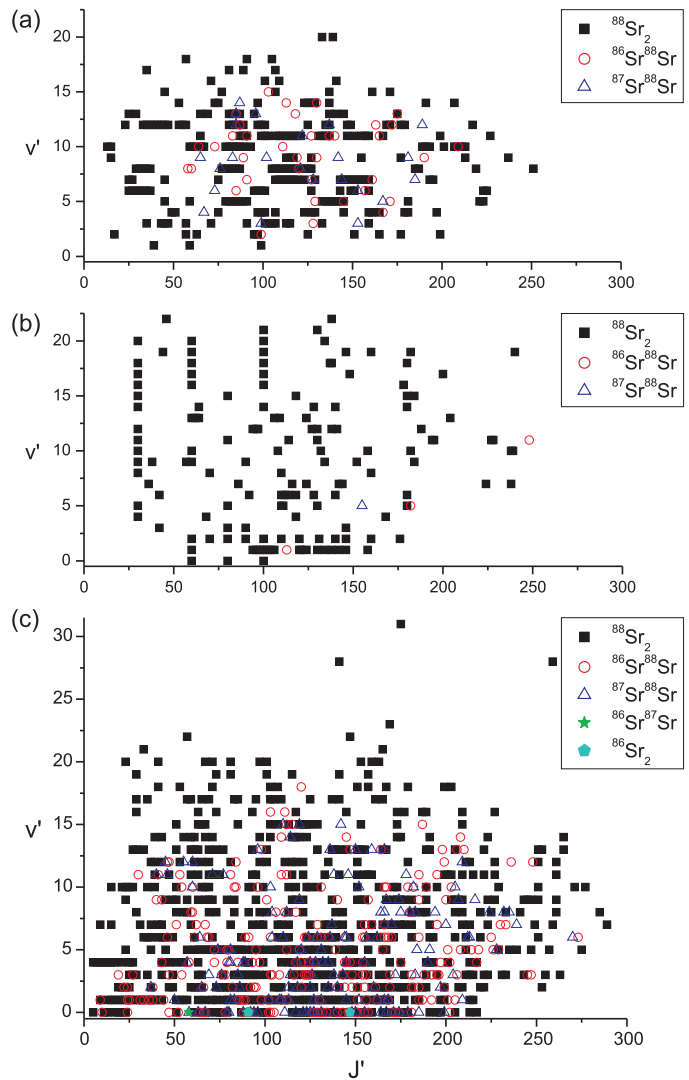}}
\caption{(Color online) Overview of the observed energy levels of state $1^1\Sigma^+_u$ (a), state $1^1\Pi_u$ (b) and state $2^1\Sigma^+_u$ (c), including all measurements of ref. \cite{SteinSr2008,SteinSrXAsymptote}.}
\label{Data}
\end{figure}

\begin{table}
\caption{Number of observed levels for the investigated electronic states for various Sr$_2$ isotopologues. The last column gives the number of ground state levels observed by fluorescence from the excited states.}
\label{Stats}
\centering
\begin{tabular}{lrcrrrr}
\hline\noalign{\smallskip}
                      & abundance &   & $1^1\Sigma^+_u$ & $1^1\Pi_u$  & $2^1\Sigma^+_u$ & X$^1\Sigma^+_g$ \\ \hline\noalign{\smallskip}
$^{84}$Sr$^{88}$Sr    &  0.92\%   &  \multirow{2}{0pt}{\huge\}} & \multirow{2}{*}{0} & \multirow{2}{*}{0} & \multirow{2}{*}{2} & \multirow{2}{*}{70} \\ 
$^{86}$Sr$_2$         &  0.97\%   &   &                 &             &                 &                 \\
$^{86}$Sr$^{87}$Sr    &  1.38\%   &   & 0               & 0           & 2               & 42              \\
$^{87}$Sr$_2$         &  0.49\%   &   & \multicolumn{4}{c}{blended by $^{86}$Sr$^{88}$Sr}                 \\
$^{86}$Sr$^{88}$Sr    & 16.28\%   &   & 37              & 3           & 223             & 4485            \\
$^{87}$Sr$^{88}$Sr    & 11.56\%   &   & 23              & 1           & 142             & 2955            \\
$^{88}$Sr$_2$         & 68.19\%   &   & 272             & 149         & 675             & 4890            \\
\hline\noalign{\smallskip}
total                 & 99.81\%   &   & 332             & 153         & 1044            & 12442           \\
\hline
\end{tabular}   
\end{table}

The data fields of the three excited states are given in figure \ref{Data} and table \ref{Stats} gives the numbers of levels for the different electronic states and isotopologues. The isotopologue $^{86}$Sr$_2$ is only observed through levels with low quantum numbers $v$ and $J$ where it cannot be distinguished from the isotopologue $^{84}$Sr$^{88}$Sr because of almost equal reduced masses. There are no independent observations of the isotopologue $^{84}$Sr$^{88}$Sr through levels with even $J'$, which do not exist for states of type $^1\Sigma^+_u$ of the homonuclear isotopologue $^{86}$Sr$_2$.  Both isotope combinations have also similar natural abundances, thus we expect that the observed progressions are the sum of contributions from both isotopologues which are excited by the same laser frequency. By the same reason the progressions from the isotopologue $^{87}$Sr$_2$ for low quantum numbers should always be blended by the stronger (factor of 33) progressions from the isotopologue $^{86}$Sr$^{88}$Sr. Thus it was never observed separately during these measurements. Of the 153 observed levels of the state $1^1\Pi_u$ only 13 are e levels, all remaining rovibrational levels are f levels. The data set for the state $2^1\Sigma^+_u$ published in \cite{SteinSr2008} consisted only of 260 levels compared to the 1044 used for the current analysis. The levels of all individual states are derived from altogether 60000 assigned lines of which 15700 were taken from our previous works \cite{SteinSr2008,SteinSrXAsymptote}. 

The assignment was done using a specially developed software as described in \cite{SteinSr2008,SteinSrXAsymptote} (in a prior version in \cite{SteinLiCs2008}). The assignment function mostly used for the current work needs as input one selected line from the loaded spectrum which is assumed to belong to a progression and the excitation frequency, where it takes as default value the frequency of the strongest line (if the user does not set another frequency). The program then runs through all $v''$ and $J''$ assignments for all possible isotopologues of all molecules of which the data of the rovibrational manifolds are loaded (for this work only data for Sr$_2$ was loaded). The first test done for each possibility of assignment is if the laser frequency fits into the progression corresponding to the quantum numbers, thus if the assumed excited level could have been excited by the laser frequency. If this is the case the spectrum is searched for additional lines which could belong to the progression. The final rating criterion for Q progressions is the highest number of additionally found 'good' lines, i.e. lines which do not overlap with other lines and have a signal-to-noise ratio better than three. In the case of a doublet progression the program selects the progression with the highest number of 'good doublets', i.e. doublets consisting of two good lines with intensity deviations from each other of less than 30\%. This selection is presented to the user, who then has to decide if he wants to save or discard the suggested assignment. This assignment function is a member of a higher order subroutine which automatically applies the function to all yet unassigned lines in the current run of evaluation. This macro routine was often run during the nights between measurement days to search for progressions overlooked by the user in the less automated assignments during the measurements. But all automatic assignment are checked in the potential fit later, which gives the stringent consistency proof of the whole dataset.

The absolute vibrational assignment of the states $1^1\Sigma^+_u$ and $1^1\Pi_u$ was first derived by the comparison of the observed relative line intensities to Franck-Condon factors calculated from preliminary RKR potentials obtained for different vibrational assignments. A vibrational assignment by counting the minima of the intensity envelopes of the observed progressions was not possible for these two states because the positions of the potential minima are too strongly shifted towards small internuclear distances compared to the ground state and thus not all minima of the intensity envelopes are visible. The vibrational assignments are later confirmed by the observation of fluorescence progressions from the weaker isotopologues with the correct frequencies predicted from mass scaling. 

The vibrational assignment of the state $2^1\Sigma^+_u$ was taken from our previous work \cite{SteinSr2008} in accord to the result in \cite{Gerber_Sr2_1984} and is later confirmed here by the methods described above. 

For all measured lines we assume an uncertainty of 0.01 cm$^{-1}$ for absolute line frequencies which equals the uncertainty of the Fourier spectrometer specified for absolute frequencies (in contrast to the specified uncertainty for frequency differences of 0.001 cm$^{-1}$). The Doppler half width for Sr$_2$ at a temperature of 1220 K is 0.02 cm$^{-1}$, but because of the weakness of the observed Sr$_2$ fluorescence it is less probable to detect progressions excited with laser frequencies in the wings of the line profiles. So the uncertainty estimation of 0.01 cm$^{-1}$ should be justified. 

\section{Results}
\label{sec:results}

As already mentioned in the above sections, to each level of the excited states belongs a fluorescence progression with individual rovibrational levels of the ground state which are employed for the assignment. The energy of the excited level is then calculated by averaging the sums of the line frequencies and the energies of the corresponding ground state levels. Thus, the level energy and the quantum  number $J'$ due to the selection rules are fixed by the very precise ground state potential \cite{SteinSrXAsymptote}. The lists of the observed excited levels can be found in the additional online material.   

The rovibrational level scheme of an electronic state with electronic quantum number $\Omega$ can be described by the conventional Dunham expansion \cite{Dunham} in many cases:

\begin{equation}
\label{GDun}
E^i_{vJ}=\sum_{k,l}Y_{lk}\left(\sqrt{\frac{\mu_0}{\mu_i}}\right)^{l+2k}{\left(v+\frac{1}{2}\right)}^l\cdot{\left[J(J+1)-\Omega^2\right]}^k \mbox{ .}
\end{equation}

The $Y_{lk}$ are the Dunham parameters of the chosen reference isotopologue $^{88}$Sr$_2$ with reduced mass $\mu_0$, while $\mu_i$ is the reduced mass of the isotopologue $i$, for which the energy ladder is considered. This simple approach is used for a first consistency check and for the convenience of predicting new transitions. Later the information will be concentrated in the derived potential energy curve of the electronic states.

\subsection{The state $1^1\Sigma^+_u$}

The reduction of all observed transitions leads to 332 different levels of state $1^1\Sigma^+_u$.

\begin{table*}
\caption{Dunham coefficients for the state 1$^1\Sigma^+_u$, for the reference isotopologue $^{88}$Sr$_2$ with $1\leq{}v'\leq10$ and $J'<220$. All values in cm$^{-1}$. The energy scale is referred to the potential minimum of the ground state.}
\label{A1Sigma_Dunham}
\centering
\begin{tabular}{rrrrr}
\hline\noalign{\smallskip}
\(l\downarrow{} k\rightarrow{}\) & 0 & 1 & 2 & 3\\ \hline\noalign{\smallskip}
0 & 12795.027 & 0.0247936 & -9.172$\times 10^{-9}$ & -3.79$\times 10^{-14}$\\
1 & 80.7129 & -1.1907$\times 10^{-4}$ & -8.2$\times 10^{-11}$ & 2.96$\times 10^{-14}$\\
2 & -0.22957 & 2.556$\times 10^{-6}$ & & -8.458$\times 10^{-15}$\\
3 & -0.01791 & -2.17$\times 10^{-7}$ & & 1.012$\times 10^{-15}$\\
4 & 1.88$\times 10^{-3}$ & & & -4.5$\times 10^{-17}$\\
5 & -8.3$\times 10^{-5}$ & & & \\
\hline
\end{tabular}
\end{table*}

Table \ref{A1Sigma_Dunham} gives a set of Dunham coefficients derived for the quantum number range $1\leq{}v'\leq10$, $J'<220$ of the state $1^1\Sigma^+_u$. Attempting to produce a set of coefficients for the full range of observation it soon became obvious that for higher $v'$ and $J'$ the levels are strongly perturbed. The value given for $Y_{00}$ is the sum of the original Dunham correction $Y_{00}$ \cite{Dunham} and the electronic term energy $T_e$ calculated with respect to the potential minimum of the ground state X$^1\Sigma^+_g$, which is 1081.64(2) cm$^{-1}$ below the asymptote $^1$S$_0 + ^1$S$_0$. The coefficient set reproduces the 196 levels within the given quantum number range with a weighted standard deviation of $\sigma=0.59$ which might indicate that the assumed measurement uncertainty of mostly 0.01 cm$^{-1}$ might be too large. The validity of the Kratzer relation (deriving the centrifugal distortion ($Y_{02}$) from the rotational ($Y_{01}$) and vibrational constant ($Y_{10}$): $Y_{02} \approx -4Y_{01}^3/Y_{10}^2$), which holds within 2\%, indicates that the minimum region of the potential is not very much affected by the perturbations. The number of digits given in table \ref{A1Sigma_Dunham} is adjusted to allow a precise reproduction of the observed energy levels, i.e. better than 0.01 cm$^{-1}$.

\begin{figure}
\resizebox{0.5\textwidth}{!}{%
\includegraphics{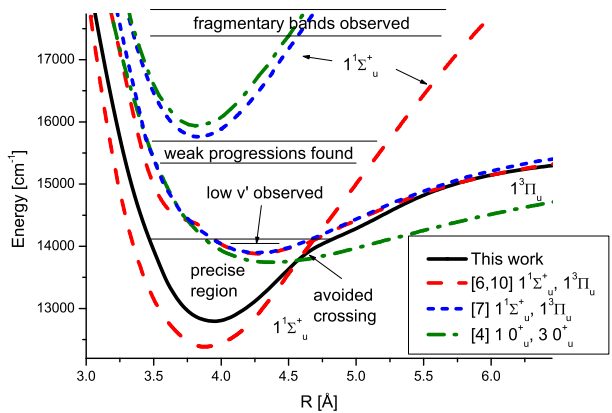}}
\caption{(Color online) Comparison of the potential of the state $1^1\Sigma^+_u$ with the \textit{ab initio} curves from Boutassetta et.al. \cite{Boutassetta,Frecon_priv}, Czuchaj et.al. \cite{Czuchaj} and from Kotochigova \cite{Kotochigova2008} calculated for the states $1^1\Sigma^+_u$ and $1^3\Pi_u$.}
\label{A1SPotComp}
\end{figure}

Since the amount of currently available data in the perturbed potential region is not sufficient for a deperturbation analysis we decided to try a single potential fit. Because of the expected appearance of an avoided crossing (see figure \ref{Overview}) a potential fit using an analytic potential representation as in \cite{SteinSrXAsymptote} will not give satisfying results when all observed levels up to $v'=20$ were included. Thus we used a more flexible potential description with cubic splines based on the algorithm of \cite{Pashov2000622} (also described in \cite{IvanovaLiRbX2010}) to fit a potential, which clearly shows the turn over to an avoided crossing reaching the energy where the \textit{ab initio} calculations \cite{Kotochigova2008,Boutassetta,Czuchaj} predict the state $1^3\Pi_u$. Obviously the adiabatic picture will become more appropriate for the current situation. According to this picture the derived potential energy curve describes the state 1 $0^+_u$, see figure \ref{A1SPotComp} (full lines), which coincides in the minimum with the state $1^1\Sigma^+_u$, but dissociates to the atomic asymptote $^3$P$_1$ + $^1$S$_0$. The central part of the potential ($R_i \leq R \leq R_a$) is described by sample points interpolated with cubic spline functions (see e.g. \cite{NumericalRecipes}) and is connected at the end points $R_i$ and $R_a$ to short and long range extensions with continuously differentiable connections to the extension functions (contrary to \cite{Pashov2000622,IvanovaLiRbX2010}). The short range extension at the inner repulsive wall ($R \leq R_i$) has the form

\begin{equation}
\label{GVi}
V_i(R)=A+\frac{B}{R^n} \mbox{ ,}
\end{equation}

\noindent and the long range part ($R \geq R_a$) is given by

\begin{equation}
\label{GVa}
V_a(R)=U_\infty-\frac{C_6}{R^6}-\frac{C_8}{R^8} \mbox{ ,}
\end{equation}

\noindent where $U_\infty$ is the asymptotic energy of the pair Sr $^3$P$_1$ + Sr $^1$P$_0$.

It is important to note that the coefficients $C_6$ and $C_8$ are only used to obtain a reasonable potential shape in the region where no or only few data are available and they should not be interpreted as long range coefficients, which would  allow conclusions on atomic properties. 

Figure \ref{A1SPotComp} shows the derived potential curve as thick full line. The position of the avoided crossing in the resulting potential, around 4.5 \AA, clearly indicates that the perturbing state is the state $1^3\Pi_{0u}$. The potential describes the data set (measurement precision of 0.01 cm$^{-1}$) with an accuracy better than a few tenth of cm$^{-1}$ in most cases. The reason for such large residuals should be related to perturbations by the spin-orbit coupled state for which at least one level is observed (likely $v'=0$ of $\Omega=0$) and by the $\Omega=1$ component of the state $1^3\Pi_u$ with its rotational coupling to $^3\Pi_{0u}$. For the potential fit the data belonging to higher lying adiabatic potentials of the coupled system marked in figure \ref{A1SPotComp} with 'low $v'$', 'weak progressions' and 'fragmentary bands' are currently not used but may be helpful for a future deperturbation analysis.

\begin{table}
\caption{Potential coefficients for the state 1$^1\Sigma^+_u$, the potential energies are calculated with respect to the potential minimum of the ground state. The given long range coefficients are only derived for a proper extrapolation to the dissociation asymptote $^1$S$_0 + ^3$P$_1$.}
\label{A1Sigma_Potential}
\centering
\begin{tabular}{lll|lll}
\hline\noalign{\smallskip}
\multicolumn{2}{l}{R [\AA]} & Energy [cm$^{-1}$] & \multicolumn{2}{l}{R [\AA]} & Energy [cm$^{-1}$]\\
\hline\noalign{\smallskip}
\multicolumn{2}{l}{3.464536} & 14161.788 & \multicolumn{2}{l}{4.209351} & 13054.842 \\
\multicolumn{2}{l}{3.546908} & 13681.495 & \multicolumn{2}{l}{4.259994} & 13141.592 \\
\multicolumn{2}{l}{3.603682} & 13407.698 & \multicolumn{2}{l}{4.331647} & 13278.570 \\
\multicolumn{2}{l}{3.635081} & 13283.387 & \multicolumn{2}{l}{4.371286} & 13359.304 \\
\multicolumn{2}{l}{3.702534} & 13068.393 & \multicolumn{2}{l}{4.446272} & 13520.281 \\
\multicolumn{2}{l}{3.766714} & 12925.453 & \multicolumn{2}{l}{4.486677} & 13607.099 \\
\multicolumn{2}{l}{3.821524} & 12843.185 & \multicolumn{2}{l}{4.542847} & 13728.271 \\
\multicolumn{2}{l}{3.886870} & 12792.860 & \multicolumn{2}{l}{4.622617} & 13873.598 \\
\multicolumn{2}{l}{4.001479} & 12821.585 & \multicolumn{2}{l}{4.753274} & 14042.967 \\
\multicolumn{2}{l}{4.061114} & 12867.134 & \multicolumn{2}{l}{4.968028} & 14256.570 \\
\multicolumn{2}{l}{4.147573} & 12963.159 & \multicolumn{2}{l}{5.119717} & 14483.744 \\
\hline
\multicolumn{3}{c|}{$R \leq R_i = 3.464536$ \AA} & 
\multicolumn{3}{c}{$R \geq R_a = 5.119717$ \AA} \\
\hline\noalign{\smallskip}
$A$ & \multicolumn{2}{l|}{10653.785 cm$^{-1}$} & $U_\infty$ & \multicolumn{2}{l}{15585.989 cm$^{-1}$} \\
$B$ & \multicolumn{2}{l|}{6066396 cm$^{-1}$\AA$^6$} & $C_6$ & \multicolumn{2}{l}{1.203457$\times 10^{7}$ cm$^{-1}$\AA$^{6}$} \\
$n$ & 6 & & $C_8$ & \multicolumn{2}{l}{2.048467$\times 10^{8}$ cm$^{-1}$\AA$^{8}$} \\
\hline
\end{tabular}
\end{table}

The potential coefficients are given in table \ref{A1Sigma_Potential}. For obtaining level energies within the lowest ten vibrational states it is much better to use the Dunham coefficients given in table \ref{A1Sigma_Dunham} or an alternative potential function contained in the online material, which describes the same reduced data set with $v' \leq 10$ as the Dunham coefficients do with a weighted standard deviation of $\sigma=0.65$.

The potential curve is certainly precise enough for a comparison with \textit{ab initio} potentials being the main goal of the present work. Figure \ref{A1SPotComp} compares graphically the resulting potential to the \textit{ab initio} calculations from Boutassetta et. al. \cite{Boutassetta,Frecon_priv} (dashed line), Czuchaj et.al. \cite{Czuchaj} (dotted) and from Kotochigova \cite{Kotochigova2008} (dash-dotted) derived for the states $1^1\Sigma^+_u$ and $1^3\Pi_u$. The potential data from \cite{Boutassetta,Frecon_priv} were originally given in hartree as energy unit and thus converted to cm$^{-1}$ for this comparison (1 hartree = 219474.63 cm$^{-1}$). We adjusted the energy offset of the complete potential set from \cite{Boutassetta,Frecon_priv} such that their ground state asymptote lies at the dissociation asymptote as given in \cite{SteinSrXAsymptote}. Usually it is more appropriate to take the potential minimum as the reference point for comparisons to \textit{ab initio} calculations, but in the current case for \cite{Czuchaj} no ground state potential is available and for \cite{Boutassetta} and \cite{Kotochigova2008} the deviations in the dissociation energies of the ground state potentials are negligible small compared to the excited states. The potential curves from \cite{Czuchaj} and \cite{Kotochigova2008} were originally referenced to the ground state asymptote and thus the energy offsets for these potentials were shifted by the experimental ground state dissociation energy reported in \cite{SteinSrXAsymptote}. In figure \ref{A1SPotComp} one can clearly see that all the potentials calculated for the state $1^3\Pi_u$ agree quite well (up to an internuclear distance of 4.5 \AA) while the potentials calculated for the state $1^1\Sigma^+_u$ show deviations as if they were different electronic states. It might be surprising, that the earliest work by Boutassetta et.al. \cite{Boutassetta} comes closest to our experimental results and overestimates the binding energy of this state by about 400 cm$^{-1}$, while the two recent works by Czuchaj et.al. \cite{Czuchaj} and Kotochigova \cite{Kotochigova2008} underestimate it by about 3000 cm$^{-1}$. For the strong deviation of the result from \cite{Czuchaj} we note, that their $1^1\Sigma^+_u$ potential looks like the $1^1\Delta_u$ potential from \cite{Boutassetta} and their $1^1\Delta_u$ potential looks like the $1^1\Sigma^+_u$ potential from \cite{Boutassetta}. If they were wrongly assigned they would only have deviations of the small magnitude as for \cite{Boutassetta}. Kotochigova \cite{Kotochigova2008} did 'full-relativistic' calculations, published adiabatic potential energy curves and wrote that 'various avoided crossings' between the low $0^+_u$ states exist. Thus it is not completely clear which of the reported potential curves should be used for this comparison. Because the published pictures suggest that all avoided crossings are at the inner repulsive potential walls, while the states look well separated and well shaped in the regions from the asymptotes to the potential minima, we took the state 3 $0^+_u$, which is the state with $\Omega=0$ from the asymptote $^1$D + $^1$S. In any case Kotochigova did not predict the observed avoided crossing between the states $1^1\Sigma^+_u$ and $1^3\Pi_{0u}$ and thus definitively the result in \cite{Kotochigova2008} shows a significant conflict with the observations.

One could ask if our assignment of the electronic state is correct and if we really observed the state $1^1\Sigma^+_u$. We observed strong line intensities and only transitions with $\Delta J=\pm 1$ and no Q lines, thus it is extremely unlikely that we observed anything different from a state of type $^1\Sigma^+_u$. The $\Omega=0^+_u$ component of a triplet state would be accessible because of the strong spin-orbit coupling of the Strontium atom but should have much lower transition dipole moments to the ground state than a $^1\Sigma^+_u$ state. Additionally, our present observations are very consistent with the experimental observations for the quite similar molecule Ca$_2$ \cite{AllardCa2CoupledStates}. 

Contrary to the calculations for the state $1^1\Sigma^+_u$ the \textit{ab initio} potentials \cite{Boutassetta,Czuchaj,Kotochigova2008} of the state $1^3\Pi_u$ will fit to the observed position of the avoided crossing between these two states. Only the long-range behavior of the result from \cite{Kotochigova2008} would correspond to an unrealistic large $C_3$ coefficient which is in contradiction to the measurement results from ultracold ensembles (e.g. \cite{ZelevinskiC2008}).

\begin{table}
\caption{Comparison of spectroscopic constants of the state $1^1\Sigma^+_u$ and the isotopologue $^{88}$Sr$_2$. The energy $T_e$ is calculated with respect to the potential minimum of the ground state, our values for $T_e$ and $R_e$ are taken from the potential fit, while the values for $\omega_e$ and $B_e$ are taken from the Dunham fit.}
\label{A1Sigma_Comp}
\centering
\begin{tabular}{lllll}
\hline\noalign{\smallskip}
Source    & $T_e$ [cm$^{-1}$] & $R_e$ [\AA{}] & $\omega_e$ [cm$^{-1}$] & $B_e$ [cm$^{-1}$] \\ \hline\noalign{\smallskip}
This work & 12796(2)          & 3.95(1)       & 80.71(3)               & 0.024794(2)       \\
\cite{Boutassetta} & 12363    & 3.850         & 79                     & 0.0259            \\
\cite{Czuchaj} & 15792        & 3.77          &                        &                   \\
\cite{Kotochigova2008} & 15940 & 3.8          &                        &                   \\
 \hline
\end{tabular}
\end{table}

Table \ref{A1Sigma_Comp} summarizes for the state $1^1\Sigma^+_u$ the experimentally determined spectroscopic constants $T_e$, equilibrium internuclear separation $R_e$, vibrational constant $\omega_e \approx Y_{10}$ and rotational constant $B_e \approx Y_{01}$ and those from theory. From \cite{Kotochigova2008} the values for the state 3 $0^+_u$ are taken. The values $D_e$ given in \cite{Czuchaj} (5440 cm$^{-1}$) and \cite{Kotochigova2008} (5292 cm$^{-1}$) are converted to $T_e$ by using the atomic level difference $^1$S$_0\longrightarrow ^1$D$_2$ (20149.700 cm$^{-1}$ \cite{SansonettiAtomicRefData}) and the $D_e$ value of the ground state $X^1\Sigma^+_g$ (1081.64(2) cm$^{-1}$ \cite{SteinSrXAsymptote}). The results for $\omega_e$ and $B_e$ from \cite{Boutassetta} show good agreement with the experimental results and deviate only by 2\% and 4\%, respectively. 

A tabulated potential function (with high density of points) derived with the data from table \ref{A1Sigma_Potential}, a highly precise potential describing only the unperturbed levels of this state in a point-wise and an analytic version (for definition see eq. (\ref{GVC}) below)  can be found in the online material.

\subsection{The state $1^1\Pi_u$}

From all fluorescence progressions we derived in total 153 levels of the state $1^1\Pi_u$.

While fitting Dunham coefficients for the state $1^1\Pi_u$ it soon became clear that this state is weakly perturbed. A small number of local perturbations is visible in the ladder of f-levels obtained from the progressions of Q lines. 

\begin{figure}
\resizebox{0.5\textwidth}{!}{%
\includegraphics{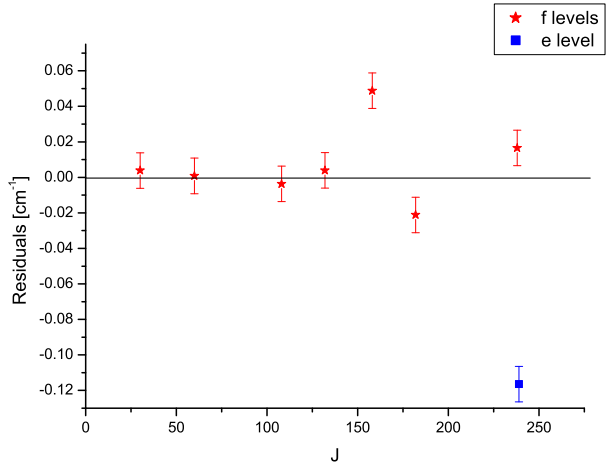}}
\caption{(Color online) Fit residuals for $v'=10$ of the state $1^1\Pi_u$ illustrating perturbations.}
\label{ResBv10}
\end{figure}

Figure \ref{ResBv10} shows the residuals for $v'=10$ of the potential fit discussed later in this subsection and gives a typical picture of the observed perturbations. Each of the eight individual levels is derived from long fluorescence progressions and the uncertainty for the energy is smaller than 0.01 cm$^{-1}$ as also the Dunham fit in the last subsection indicated. Most of the f-levels shown are not significantly perturbed while there is one weak but significant local perturbation visible between $J'=158$ and $J'=182$. The only e-level observed for this $v'$ is more strongly shifted by a local perturbation caused by $v'=2$ of the state $2^1\Sigma^+_u$ as one can see using the data of section \ref{sec:state2A1S}. 

Theoretically such perturbations of f-levels can only be caused by singlet states of type $^1\Sigma^-_u$, $^1\Pi_u$, $^1\Delta_u$ or triplet states with u-symmetry. Inspecting the \textit{ab initio} calculations \cite{Boutassetta} the state $^1\Delta_u$ from the same asymptote $^1$D + $^1$S seems to be a good candidate as perturber, since it will have by far the largest Franck-Condon overlap with the state $1^1\Pi_u$. 

\begin{table}
\caption{Dunham coefficients $Y_{lk}$ for the f-levels of the state 1$^1\Pi_u$ and the reference isotopologue $^{88}$Sr$_2$. All values in cm$^{-1}$. $Y_{00}$ is calculated with respect to the potential minimum of the ground state and includes the electronic term energy $T_e$.}
\label{B1Pi_Dunham}
\centering
\begin{tabular}{rrrrr}
\hline\noalign{\smallskip}
\(l\downarrow{} k\rightarrow{}\) & 0 & 1 & 2\\ \hline\noalign{\smallskip}
0 & 16617.8249 & 0.02341446 & -6.943$\times 10^{-9}$\\
1 & 86.2991 & -6.272$\times 10^{-5}$ & \\
2 & -0.21086 & -9.7$\times 10^{-8}$ & -3$\times 10^{-12}$\\
3 & -2.3$\times 10^{-5}$ & & \\
4 & -2.42$\times 10^{-5}$ & &\\
\hline
\end{tabular}
\end{table}

Table \ref{B1Pi_Dunham} gives the resulting Dunham coefficients which can be used to reproduce the term energies of the unperturbed levels from $v'=0$ to $v'=22$ according to eq. (\ref{GDun}). 32 of the 153 observed levels, including all e levels, were removed from the Dunham fit because of perturbations. The weighted standard deviation of this fit is $\sigma=0.85$.

Though all observed e-levels with rotational quantum number $J'>180$ are systematically shifted to lower energies than extrapolated from the observed f-levels, a consolidated description of the $\lambda$-doubling can only be derived by a full deperturbation analysis of the system. The corresponding levels of the state $2^1\Sigma^+_u$ are shifted just into the opposite direction as the e-levels of the state $1^1\Pi_u$. In the energy region, where for both states a sufficiently dense data set exists and precise frequencies can be calculated from the potentials, the extrapolated energies of corresponding bands of both states reach the same values in the rotational ladder around $J' = 250$. The e-levels are additionally perturbed by the state $2^1\Sigma^+_u$, but they could also be perturbed by highly excited vibrational levels of the state $1^1\Sigma^+_u$ evaluated in the previous subsection.   

The Dunham coefficients were used to create a RKR potential which then served as a starting potential for the fit of an analytical potential representation, as defined in \cite{SteinSrXAsymptote} for the ground state. The central part of this potential ($R_i\leq R\leq R_a$) is represented as

\begin{equation}
\label{GVC}
V_c(R)=T_m+\sum_{j\geq1}a_jx^j
\end{equation} 

\noindent with the nonlinear mapping function

\begin{equation}
\label{GVx}
x=\frac{R-R_m}{R+bR_m} \mbox{ .}
\end{equation}

The inner repulsive wall ($R < R_i$) has the same form as for the potential used for the state $1^1\Sigma^+_u$ and is given in eq. (\ref{GVi}) and similarly for the long range part ($R > R_a$) given in eq. (\ref{GVa}). Here the $a_j$ and $T_m$ are fit parameters while $b$, $R_i$, $R_m$, $R_a$ and $n$ are manually chosen to optimize the result. $A$ and $B$ from eq. (\ref{GVi}) and the $C_i$ from eq. (\ref{GVa}) are adjusted with $i=8$ and 10 to get continuously differentiable connections at the points $R_i$ and $R_a$, respectively. $U_\infty$ is calculated by adding to the difference of the atomic levels $^1$S$_0$ -- $^1$D$_2$ of 20149.700 cm$^{-1}$ \cite{SansonettiAtomicRefData} the ground state dissociation energy of 1081.64(2) cm$^{-1}$ \cite{SteinSrXAsymptote}. The long range connection is mainly done to show the proper correlation of this state to the atomic asymptote. 

\begin{figure}
\resizebox{0.5\textwidth}{!}{%
\includegraphics{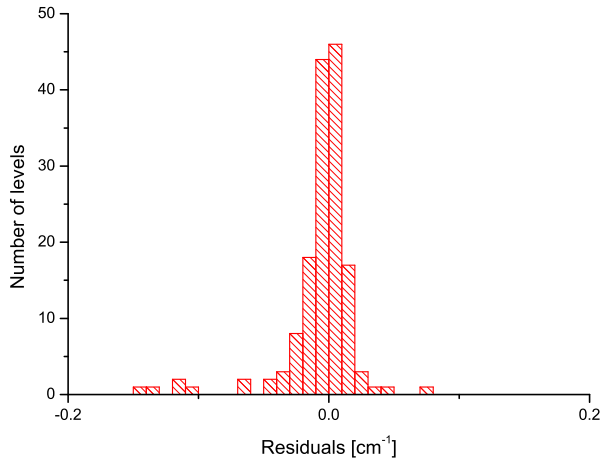}}
\caption{(Color online) Residuals of the potential fit of the state $1^1\Pi_u$ plotted as a histogram. Few perturbed levels showing deviations of up to 1 cm$^{-1}$ are out of scale in this figure.}
\label{ResB}
\end{figure}

For the potential fit the same data set was used as for the Dunham fit. Figure \ref{ResB} shows a histogram of the residuals of the potential fit of the levels assumed as being unperturbed. The weighted standard deviation of the potential fit is $\sigma=0.86$.  One can clearly see that the overwhelming amount of levels is centered around the middle with a full width at half maximum of less than 0.02 cm$^{-1} = 2\sigma$ and only few perturbed levels show larger deviations. 

\begin{table}
\caption{Potential coefficient for the state $1^1\Pi_u$. The energy is calculated with respect to the potential minimum of the ground state. The coefficients $C_8$ and $C_{10}$ are simply used as continuously differentiable extensions from the deeply bound potential region to the atomic asymptote $^1$S$_0$ + $^1$D$_2$.}
\label{B1Pi_uPotential}
\centering
\begin{tabular}{lp{0.1\textwidth}r}\hline
$a_{1}$   && -5.715$\times 10^{-1}$ cm$^{-1}$ \\
$a_{2}$   && 1.6825387$\times 10^{4}$ cm$^{-1}$ \\
$a_{3}$   && 1.320905$\times 10^{4}$ cm$^{-1}$ \\
$a_{4}$   && -1.03130$\times 10^{3}$ cm$^{-1}$ \\
$a_{5}$   && -9.40946$\times 10^{3}$ cm$^{-1}$ \\
$a_{6}$   && -1.92645$\times 10^{4}$ cm$^{-1}$ \\
$a_{7}$   && -7.423611$\times 10^{4}$ cm$^{-1}$ \\
$a_{8}$   && -1.44756$\times 10^{5}$ cm$^{-1}$ \\
\hline
$b$     && -0.54    \\
$R_m$ && 4.047308 \AA \\ 
$T_m$ && 16617.8639 cm$^{-1}$ \\ \hline
$R_i$ && 3.524 \AA \\
$n$     && 6    \\
$A$     && 1.4062979$\times 10^{4}$ cm$^{-1}$ \\
$B$     && 8.341380$\times 10^{6}$ cm$^{-1}$\AA$^n$ \\ \hline
$R_a$ && 4.88 \AA \\
$C_{8}$ && 1.8205183$\times 10^{9}$ cm$^{-1}$\AA$^{8}$ \\
$C_{10}$ && -2.266162$\times 10^{10}$ cm$^{-1}$\AA$^{10}$ \\
$U_\infty$ && 21231.34(2) cm$^{-1}$ \\ \hline
\end{tabular}
\end{table}

In table \ref{B1Pi_uPotential} the resulting potential coefficients are listed which can be used to construct the potential according to eqs. (\ref{GVi}-\ref{GVx}). 

\begin{figure}
\resizebox{0.5\textwidth}{!}{%
\includegraphics{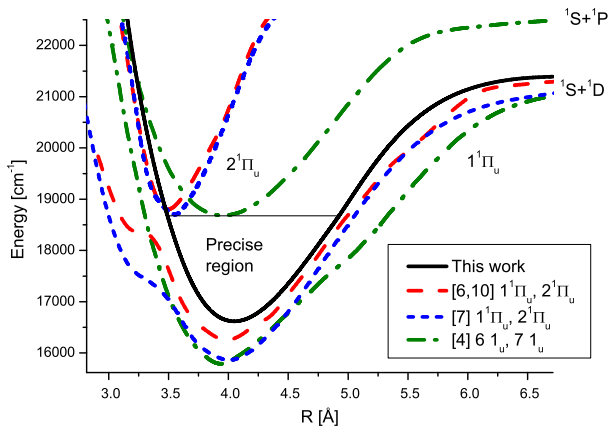}}
\caption{(Color online) Comparison of the potential energy curve with \textit{ab initio} calculations of the lowest two $^1\Pi_u$ states from Boutassetta et.al. \cite{Boutassetta,Frecon_priv}, Czuchaj et.al. \cite{Czuchaj} ($1^1\Pi_u$ and $2^1\Pi_u$) and Kotochigova \cite{Kotochigova2008} (6(1$_u$) and 7(1$_u$)). The energies are calculated with respect to the potential minimum of the ground state. The region with precise spectroscopic data is indicated.}
\label{CompB}
\end{figure}

\begin{table*}
\caption{Comparison of spectroscopic constants of the state $1^1\Pi_u$, the energy $T_e$ refers to the potential minimum of the ground state.}
\label{B1Pi_Comp}
\centering
\begin{tabular}{lllll}
\hline\noalign{\smallskip}
Source    & $T_e$ [cm$^{-1}$] & $R_e$ [\AA{}] & $\omega_e$ [cm$^{-1}$] & $B_e$ [cm$^{-1}$] \\ \hline\noalign{\smallskip}
This work & 16617.86(2)       & 4.0473(2)     & 86.300(3)              & 0.023415(2)       \\
\cite{Boutassetta} & 16243    & 3.952         & 96                     & 0.0246            \\
\cite{Czuchaj} & 15821        & 3.97          &                        &                   \\
\cite{Kotochigova2008} & 15818 & 3.9          &                        &                   \\
 \hline
\end{tabular}
\end{table*}

A comparison of the potential from this work with the \textit{ab initio} results is shown in figure \ref{CompB}. Here it is clearly visible that all calculations significantly overestimate the binding energy (i.e. too low $T_e$ values) of this state and again the earliest work \cite{Boutassetta} shows by far the best agreement, a deviation of 375 cm$^{-1}$, while the more recent works \cite{Czuchaj,Kotochigova2008} show deviations of about 800 cm$^{-1}$. Table \ref{B1Pi_Comp} gives a comparison of the spectroscopic constants showing significant deviations. Looking into the potential plot given in Fig. 2b of \cite{Kotochigova2008} one can see that the adiabatic potential curve (state \mbox{6 1$_u$}) has an avoided crossing with a $\Omega=1_u$ component of a triplet state ($^3$D + $^1$S) at 5 \AA{} and thus the diabatic potential curve of the state $1^1\Pi_u$ to $^1$D + $^1$S would even deviate more strongly. The pronounced avoided crossing at the inner potential wall predicted by \cite{Boutassetta} and \cite{Kotochigova2008} is also not observed, but if the binding energy of the state $2^1\Pi_u$ by \cite{Boutassetta} is only little more overestimated than the binding energy of the state $1^1\Pi_u$ it would be outside the investigated energy region.  

A point-wise potential of the state $1^1\Pi_u$ is given in the additional online material.

\subsection{The state 2(A)$^1\Sigma^+_u$}
\label{sec:state2A1S}

\begin{table*}
\caption{Dunham coefficients for the unperturbed levels up to $v'=23$ of the state 2(A)$^1\Sigma^+_u$ for the reference isotopologue $^{88}$Sr$_2$. All values in cm$^{-1}$. The energy refers to the minimum of the ground state potential.}
\label{C1Sigma_Dunham}
\centering
\begin{tabular}{rrrrr}
\hline\noalign{\smallskip}
\(l\downarrow{} k\rightarrow{}\) & 0 & 1          & 2                     & 3                     \\ \hline\noalign{\smallskip}
0 & 17358.7246           & 0.0219695              & -5.98$\times 10^{-9}$ &                       \\
1 & 84.2157              & -6.787$\times 10^{-5}$ & -4$\times 10^{-11}$   & -1.4$\times 10^{-16}$ \\
2 & -0.2665              & -4.82$\times 10^{-7}$  & -1.7$\times 10^{-12}$ &                       \\
3 & -0.00113             & -7$\times 10^{-9}$     &                       &                       \\
4 & -1.1$\times 10^{-5}$ &                        &                       &                       \\
\hline
\end{tabular}
\end{table*}

For state 2(A)$^1\Sigma^+_u$ we collected the largest set of data, namely 1044 levels. Table \ref{C1Sigma_Dunham} gives Dunham coefficients. They can be used according to eq. (\ref{GDun}) to calculate the term energies of the unperturbed levels of all isotopologues up to $v'=23$. The Dunham coefficients reproduce with a weighted standard deviation of $\sigma=0.78$ the 1029 observed levels in the range up to $v'=23$ except 20 significantly perturbed ones. 

\begin{figure}
\resizebox{0.5\textwidth}{!}{%
\includegraphics{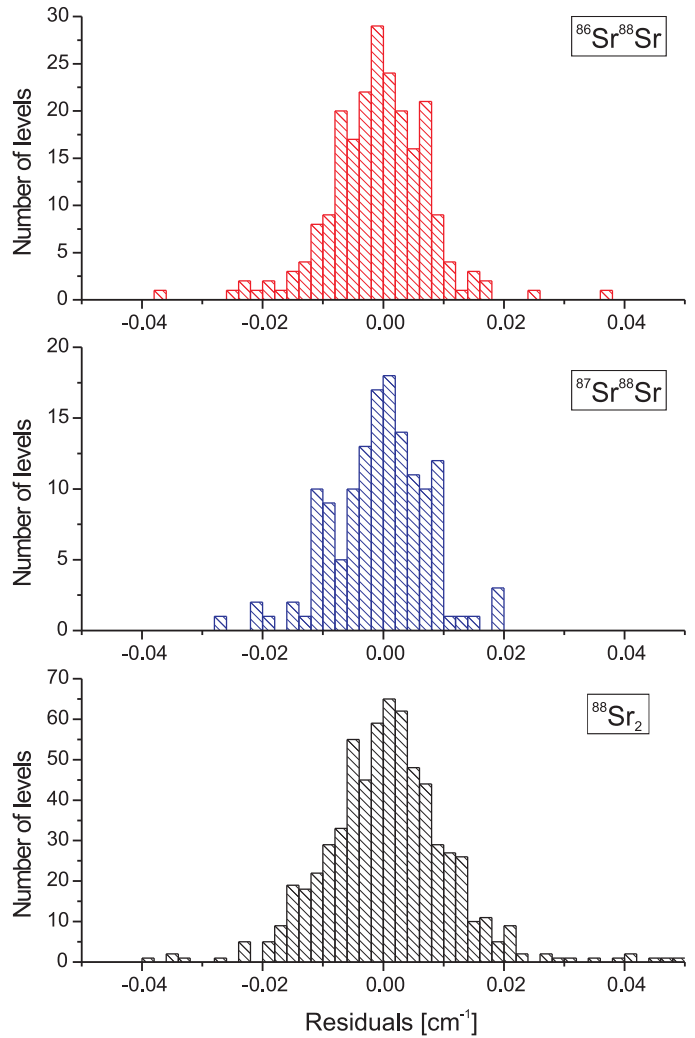}}
\caption{(Color online) Histograms of the residuals of the potential fit for the state 2(A)$^1\Sigma^+_u$ for the three most abundant isotopologues $^{86}$Sr$^{88}$Sr, $^{87}$Sr$^{88}$Sr and $^{88}$Sr$_2$.}
\label{ResC}
\end{figure}

Compared to our earlier result \cite{SteinSr2008} where only data up to $v'=12$ were available the potential function could be significantly extended to the range from 3.7 to 5.3 \AA. Additionally to the 1009 levels used for the Dunham fit, 3 levels with $v'=28$ and $v'=31$ excited with the 514 nm line of the Ar$^+$ laser are included in the potential fit, of which one of the two levels with $v'=28$ had to be removed again, likely because of a perturbation. The remaining 12 of the observed 1044 levels belong to higher vibrational levels and are not yet assigned to a $v'$ quantum number. Histograms of the residuals of the fit are given in figure \ref{ResC} for the three most abundant isotopologues $^{86}$Sr$^{88}$Sr, $^{87}$Sr$^{88}$Sr and $^{88}$Sr$_2$. A few perturbed levels showing deviations up to 0.6 cm$^{-1}$ are not included. One can clearly see that the unperturbed or weakly perturbed levels are centered around zero with a width at half maximum of less than $\pm 0.01$ cm$^{-1}$ which is the estimated measurement uncertainty for most levels and correlates to a weighted standard deviation of the fit of the 1011 weakly perturbed levels of $\sigma=0.88$. The few levels of the weak isotopologues $^{86}$Sr$_2$ ($^{84}$Sr$^{88}$Sr) and $^{86}$Sr$^{87}$Sr show deviations from the predictions by the potential of less than 0.01 cm$^{-1}$, a quite satisfactory situation. 

\begin{figure}
\resizebox{0.5\textwidth}{!}{%
\includegraphics{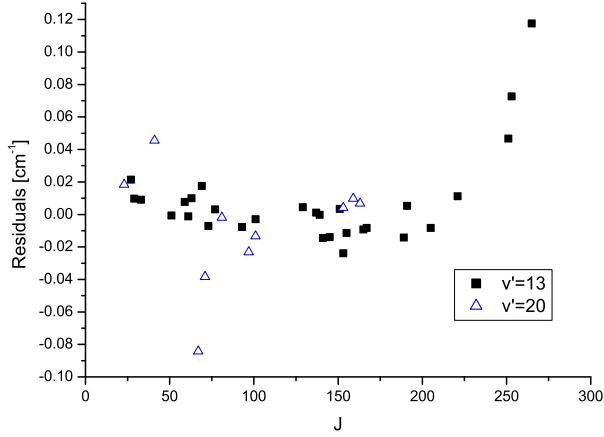}}
\caption{(Color online) The residuals of the 2(A)$^1\Sigma^+_u$ potential plotted in dependency of $J'$ for $v'=13$ and $v'=20$ of the reference isotopologue $^{88}$Sr$_2$ to illustrate perturbations between $J'=41$ and 67 and $J'>250$.}
\label{ResCBJ}
\end{figure} 

Figure \ref{ResCBJ} illustrates observed perturbations by displaying residuals as functions of $J'$ for the vibrational levels $v'=13$ and $v'=20$ of the reference isotopologue $^{88}$Sr$_2$. For $v'=13$ one can see a fast growing deviation for $J'>200$ indicating a perturbation at very high $J'$. By comparing calculations for level energies of $v'=13$ with those for the state $1^1\Pi_u$ one finds that the $v'=21$ stack of this state crosses the one of the state $2^1\Sigma^+_u$ coming from smaller energies at $J'=271$. The highest observed level of $v'=13$ of $2^1\Sigma^+_u$ corresponds to $J'=265$. Such perturbations at high $J'$ (moving slowly to lower $J'$ for growing $v'$) are observed for all $v'$ of state $2^1\Sigma^+_u$ for which a sufficient amount of data is available. For $v'=20$ one sees in figure \ref{ResCBJ} a perturbation between $J'=41$ and $J'=67$, which could stem from $v'=29$ of state $1^1\Pi_u$, but such extrapolation out of the known range of $1^1\Pi_u$ is not yet reliable. 

\begin{table}
\caption{Potential coefficients for the state 2(A)$^1\Sigma^+_u$. The energy is referenced to the potential minimum of the ground state. The long range coefficient $C_3$ is the experimental value from \cite{Yasuda}, while the coefficients $C_8$ and $C_{10}$ are used for a continuously differentiable connection at the point $R_a$.}
\label{C1Sigma+uPotential}
\centering
\begin{tabular}{lp{0.1\textwidth}r}\hline \noalign{\smallskip}
$a_{1}$   && 3.89$\times 10^{-2}$ cm$^{-1}$ \\
$a_{2}$   && 3.6222297$\times 10^{4}$ cm$^{-1}$ \\
$a_{3}$   && 7.47647$\times 10^{2}$ cm$^{-1}$ \\
$a_{4}$   && -5.9774017$\times 10^{4}$ cm$^{-1}$ \\
$a_{5}$   && -1.2795502$\times 10^{5}$ cm$^{-1}$ \\
$a_{6}$   && -8.192268$\times 10^{4}$ cm$^{-1}$ \\
$a_{7}$   && 3.309500$\times 10^{5}$ cm$^{-1}$ \\
$a_{8}$   && 8.88141$\times 10^{4}$ cm$^{-1}$ \\
$a_{9}$   && 4.626819$\times 10^{5}$ cm$^{-1}$ \\\hline \noalign{\smallskip}
$b$     && -0.33    \\
$R_m$ && 4.17828223 \AA \\ 
$T_m$ && 17358.7496 cm$^{-1}$ \\ \hline \noalign{\smallskip}
$R_i$ && 3.49 \AA \\
$n$     && 6    \\
$A$     && 1.4357908$\times 10^{4}$ cm$^{-1}$ \\
$B$     && 1.153771543$\times 10^{7}$ cm$^{-1}$\AA$^n$ \\ \hline \noalign{\smallskip}
$R_a$ && 6.1 \AA \\
$C_{3}$ && 5.9712(42)$\times 10^{5}$ cm$^{-1}$\AA$^{3}$ \cite{Yasuda}\\
$C_{8}$ && -2.7292900$\times 10^{9}$ cm$^{-1}$\AA$^{8}$ \\
$C_{10}$ && 5.5935384$\times 10^{10}$ cm$^{-1}$\AA$^{10}$ \\
$U_\infty$ && 22780.122(20) cm$^{-1}$ \\ \hline
\end{tabular}
\end{table}

In table \ref{C1Sigma+uPotential} the potential coefficients for the state 2(A)$^1\Sigma^+_u$ are collected, which can be used to calculate the potential energy curve by using the eqs. (\ref{GVi}-\ref{GVx}). Here the long range coefficient $C_3$ is taken from the photoassociation work \cite{Yasuda}, while the coefficients $C_8$ and $C_{10}$ are simply derived for producing a continuously differentiable connection to the inner part of the potential at the point $R_a$ and to bridge the gap between this point (at 6.1 \AA) and the real long range region. $U_\infty$ is calculated by adding to the frequency of the atomic transition $^1$S$_0 \longrightarrow ^1$P$_1$ of 21698.482 cm$^{-1}$ \cite{SansonettiAtomicRefData} the ground state dissociation energy of 1081.64(2) cm$^{-1}$ \cite{SteinSrXAsymptote}.

\begin{figure}
\resizebox{0.5\textwidth}{!}{%
\includegraphics{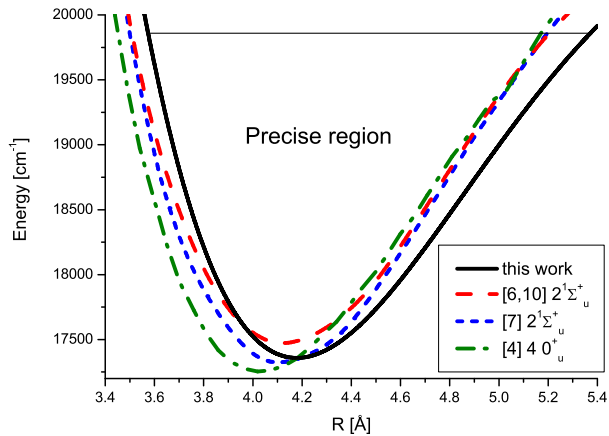}}
\caption{(Color online) Comparison of the potential for the state 2(A)$^1\Sigma^+_u$ with the \textit{ab initio} calculations from Boutassetta et.al. \cite{Boutassetta}, Czuchaj et.al. \cite{Czuchaj}, and Kotochigova \cite{Kotochigova2008}.}
\label{C1SComp}
\end{figure} 

\begin{table*}
\caption{Comparison of the spectroscopic constants of the state 2(A)$^1\Sigma^+_u$ with \textit{ab initio} calculations \cite{Boutassetta,Czuchaj,Kotochigova2008} and an earlier experimental work \cite{Gerber_Sr2_1984}. The energy $T_e$ refers to the potential minimum of the ground state.}
\label{C1Sigma_Comp}
\centering
\begin{tabular}{lllll}
\hline\noalign{\smallskip}
Source    & $T_e$ [cm$^{-1}$] & $R_e$ [\AA{}] & $\omega_e$ [cm$^{-1}$] & $B_e$ [cm$^{-1}$] \\ \hline\noalign{\smallskip}
This work & 17358.75(1)       & 4.1783(1)     & 84.215(1)              & 0.021969(1)       \\
\cite{Gerber_Sr2_1984} & 17357.9(2) & 3.952   & 85.07(34)              & 0.02456(5)        \\
\cite{Boutassetta} & 17541    & 4.099         & 83                     & 0.0229            \\
\cite{Czuchaj} & 17290        & 4.10          &                        &                   \\
\cite{Kotochigova2008} & 17269 & 4.03         & 88                     &                   \\
 \hline
\end{tabular}
\end{table*}

Figure \ref{C1SComp} shows a comparison of the resulting potential to the \textit{ab initio} calculations and table \ref{C1Sigma_Comp} summarizes the spectroscopic constants. For this electronic state the deviations are significantly smaller compared to the two other studied electronic states. Here the result from Czuchaj et.al. \cite{Czuchaj} shows the best agreement and overestimates the dissociation energy by only 69 cm$^{-1}$ and the calculation by Boutassetta et.al. \cite{Boutassetta} contrary to the other states gives the largest deviations by underestimating the binding energy by 182 cm$^{-1}$. Table \ref{C1Sigma_Comp} indicates significant deviations of the present work to an earlier experimental work \cite{Gerber_Sr2_1984} mainly for the rotational constant. They are due to a wrong rotational assignment which was already pointed out in our earlier publication \cite{SteinSr2008}.

A point-wise version of the potential and a list of the energy levels used for the potential fit can be found in the additional online material. 

\section{Conclusion}
\label{sec:conclusion}

In this work we present a potential energy curve around the potential minimum for the state $1^1\Sigma^+_u$ of which only little was known from matrix spectroscopy \cite{Miller1977,Miller1978,Miller1980}, a precise potential for the minimum of the weakly perturbed state $1^1\Pi_u$, which was completely unknown experimentally, and a substantially improved potential energy curve compared to our earlier work \cite{SteinSr2008} for the state $2^1\Sigma^+_u$. The results are compared to the currently available \textit{ab initio} calculations \cite{Kotochigova2008,Boutassetta,Czuchaj} which deviate strongly in several cases from our potentials and among each other. 

The binding energy of the state $1^1\Sigma^+_u$ is not very well predicted by all available calculations, the deviations reach from an overestimation of the binding energy by about 400 cm$^{-1}$ in \cite{Boutassetta} to an underestimation by 3000 cm$^{-1}$ in \cite{Czuchaj,Kotochigova2008}. The presented potential is determined by experimental data in the range of $v'=1$ to $v'=20$ which corresponds to internuclear distances from 3.5 \AA{} to 5.0 \AA{} and an energy interval of 1500 cm$^{-1}$. The potential function of the state $1^1\Sigma^+_u$ shows a bent like an avoided crossing which corresponds to the similar case in Ca$_2$ \cite{AllardCa2CoupledStates} where the coupling between $^1\Sigma^+_u$ and $^3\Pi_u$ was studied. Thus the potential presented in table \ref{A1Sigma_Potential} should represent the lower adiabatic component $\Omega = 0^+_u$ of the coupled system of these states. Surprisingly, all \textit{ab initio} calculations give the desired state $^3\Pi_u$ just around the energy where we find the bent in the potential curve. But Czuchaj et.al. \cite{Czuchaj} and Kotochigova \cite{Kotochigova2008} failed to predict the coupling between $1^1\Sigma^+_u$ and $1^3\Pi_u$ because the crossing of these states is high up in the repulsive branch. The calculation from Boutassetta et.al. \cite{Boutassetta} put the crossing at the right ballpark, but the precision is far not sufficient to use their calculation for a start of a deperturbation analysis. We will collect more data in the region where the strong coupling is expected for making a quantitative analysis of the coupled system $1^1\Sigma^+_u + 1^3\Pi_u$. 

The state $1^1\Pi_u$ is overestimated by all calculations in its binding energy, where \cite{Boutassetta} shows the closest result deviating by 375 cm$^{-1}$, while the recent works \cite{Czuchaj,Kotochigova2008} deviate by almost 800 cm$^{-1}$. Our experimental potential energy curve is determined by observed levels up to $v'=22$ and covers the region of internuclear distances of 3.5 \AA{} to 4.9 \AA{} and an energy up to 1800 cm$^{-1}$ above the minimum. A strongly pronounced avoided crossing at the inner repulsive potential wall with the state $2^1\Pi_u$, predicted by \cite{Boutassetta,Czuchaj} within the observed energy region, is not observed. However, if the binding energy of the state $2^1\Pi_u$ were overestimated a bit more than the binding energy of the state $1^1\Pi_u$ in their calculations the avoided crossing would move outside the observation region. 

The electronic state best described by the theoretical calculations is the state 2(A)$^1\Sigma^+_u$, where all calculations find a binding energy deviating not more than 200 cm$^{-1}$ from our spectroscopic result. \cite{Czuchaj} shows the best agreement by overestimating the binding energy by only 62 cm$^{-1}$. However there are significant shifts in the internuclear distances. The potential energy curve we present for this state is constructed from our data including vibrational levels from $v'=0$ to $v'=31$ in an interval of internuclear distances from 3.7 \AA{} to 5.3 \AA{} and up to 2400 cm$^{-1}$ above the potential minimum.

In summary, for the currently available \textit{ab initio} calculations \cite{Kotochigova2008,Boutassetta,Czuchaj} we found that surprisingly the earliest results \cite{Boutassetta} are the most useful hints for our spectroscopic work and their potential curves were partly used during this work for first calculations of Franck-Condon factors and as a guide for choosing the excitation regions of a selected electronic state. Nevertheless, all calculations are far off from being applicable for direct calculations of excitation frequencies for experiments with cold molecules. As already mentioned for Ca$_2$ \cite{AllardCa2CoupledStates} we found surprisingly large deviations in the electronic structure calculated in \textit{ab initio} work for Sr$_2$, too, and hope that our results might stimulate new efforts of the theorists.

What can be concluded from our investigations regarding production of cold molecules from ensembles of cold atoms? By far the level $v'=4$ of the state 2(A)$^1\Sigma^+_u$ has the best Franck-Condon overlap with ground state levels close to the asymptote and deeply bound ones. No other excited level of the investigated states is better suited than the one used in \cite{SteinSrXAsymptote} for the investigation of the ground state asymptote. 

In \cite{Zelevinsky2008,ZelevinskiC2008,KotochigovaProp2009} it is proposed, based on the calculations in \cite{Kotochigova2008}, to use levels of the state $1^3\Pi_{0u}$ (1 $0^+_u$) as intermediate levels for a coherent transfer process of ultracold molecules from near asymptotic ground state levels to deeply bound levels of the ground state. At this moment we can already state that the potential energy curve of 1 $0^+_u$ looks completely different than the one calculated in \cite{Kotochigova2008}. If there are nevertheless rovibrational levels of the state $1^3\Pi_u$ which are well suited for the proposed experiments we cannot judge at the moment since we did not reach that proposed frequency region. But we will try to answer this question when we continue on the coupled system of $1^1\Sigma^+_u$ -- $1^3\Pi_{0u}$ in the next future similarly to our study on Ca$_2$ \cite{AllardCa2CoupledStates}. 

\section{Acknowledgments}

This work was supported by the Deutsche Forschungsgemeinschaft within the Sonderforschungsbereich SFB 407 and the cluster of excellence QUEST.

\bibliography{Sr2_excited}

\begin{thebibliography}{23}

\bibitem{Zelevinsky2008}
T.~Zelevinsky, S.~Kotochigova, J.~Ye, Phys. Rev. Lett. \textbf{100}, 043201
  (2008)

\bibitem{ZelevinskiC2008}
T.~Zelevinsky, S.~Blatt, M.M. Boyd, G.K. Campbell, A.D. Ludlow, J.~Ye, Chem.
  Phys. Chem. \textbf{9}(3), 375 (2008)

\bibitem{KotochigovaProp2009}
S.~Kotochigova, T.~Zelevinsky, J.~Ye, Phys. Rev. A \textbf{79}(1), 012504
  (2009)

\bibitem{Kotochigova2008}
S.~Kotochigova, J. Chem. Phys. \textbf{128}, 024303 (2008)

\bibitem{AllardCa2CoupledStates}
O.~Allard, S.~Falke, A.~Pashov, O.~Dulieu, H.~Kn\"ockel, E.~Tiemann, Eur. Phys.
  J. D \textbf{35}, 483 (2005)

\bibitem{Boutassetta}
N.~Boutassetta, A.R. Allouche, M.~Aubert-Fr\'econ, Phys. Rev. A \textbf{53},
  3845 (1996)

\bibitem{Czuchaj}
E.~Czuchaj, M.~Kro\'snicki, H.~Stoll, Chem. Phys. Lett. \textbf{371}, 401
  (2003)

\bibitem{SteinSr2008}
A.~Stein, H.~Kn\"ockel, E.~Tiemann, Phys. Rev. A \textbf{78}, 042508 (2008)

\bibitem{SteinSrXAsymptote}
A.~Stein, H.~Kn\"ockel, E.~Tiemann, Eur. Phys. J. D \textbf{57}(2), 171 (2010)

\bibitem{Frecon_priv}
Listing of potential energy curves were kindly provided by M. Aubert-Fr\'econ,
  private communication (2006)

\bibitem{Miller1977}
J.C. Miller, B.S. Ault, L.~Andrews, J. Chem. Phys. \textbf{67}, 2478 (1977)

\bibitem{Miller1978}
J.C. Miller, L.~Andrews, J. Chem. Phys. \textbf{69}, 936 (1978)

\bibitem{Miller1980}
J.C. Miller, L.~Andrews, Appl. Spectrosc. Rev. \textbf{16}, 1 (1980)

\bibitem{Bergemann1980}
T.~Bergemann, P.F. Liao, J. Chem. Phys. \textbf{72}, 886 (1980)

\bibitem{Gerber_Sr2_1984}
G.~Gerber, R.~M\"oller, H.~Schneider, J. Chem. Phys. \textbf{81}, 1538 (1984)

\bibitem{SchneiderDip}
H. Schneider, Diploma thesis, Universit\"at Freiburg (1984)

\bibitem{SteinLiCs2008}
A.~Stein, A.~Pashov, P.F. Staanum, H.~Kn\"ockel, E.~Tiemann, Eur. Phys. J. D
  \textbf{48}, 177 (2008)

\bibitem{Dunham}
J.L. Dunham, Phys. Rev. \textbf{41}(6), 721 (1932)

\bibitem{Pashov2000622}
A.~Pashov, W.~Jastrzebski, P.~Kowalczyk, Computer Physics Communications
  \textbf{128}(3), 622  (2000), ISSN 0010-4655

\bibitem{IvanovaLiRbX2010}
M.~Ivanova, A.~Stein, A.~Pashov, H.~Kn\"ockel, E.~Tiemann, J. Chem. Phys.
  \textbf{134}, 024321 (2011)

\bibitem{NumericalRecipes}
W.H. Press, S.A. Teukolsky, W.T. Vetterling, B.T. Flannery, \emph{Numerical
  Recipes, Third Edition} (Cambridge University Press, New York, 2007)

\bibitem{SansonettiAtomicRefData}
J.E. Sansonetti, W.C. Martin, J. Phys. Chem. Ref. Data \textbf{34}, 1559 (2005)

\bibitem{Yasuda}
M.~Yasuda, T.~Kishimoto, M.~Takamoto, H.~Katori, Phys. Rev. A \textbf{73},
  011403(R) (2006)

\end{thebibliography}

\end{document}